\documentclass[12pt]{article}
\usepackage{graphicx}
\def\mydate{December 1, 2004}
\def\ignore#1{{}}

\newcounter{sxn}

\newcounter{axn}

\date{}

\newdimen\mybaselineskip
\renewcommand{\baselinestretch}{1.25}

\newcommand{\beeq}{\begin{equation}}
\newcommand{\eneq}{\end{equation}}
\newcommand{\beqn}{\begin{eqnarray}}
\newcommand{\eeqn}{\end{eqnarray}}

\def\mybig{\displaystyle \strut }

\def\dd{\partial}
\def\la{\raise.16ex\hbox{$\langle$}\lower.16ex\hbox{}  }
\def\ra{\, \raise.16ex\hbox{$\rangle$}\lower.16ex\hbox{} }
\def\go{\rightarrow}

\def\onehalf{ \hbox{${1\over 2}$} }

\def\Tr{{\rm Tr \,}}
\def\tr{{\rm tr \,}}
\def\eff{{\rm eff}}

\def\Ithree{{\bf 1}_{3\times 3}}

\def\ep{\epsilon}
\def\psibar{ \psi \kern-.65em\raise.6em\hbox{$-$} }
\def\psibarl{ \psi \kern-.65em\raise.6em\hbox{$-$} \lower.6em\hbox{} }

\def\myfrac#1#2{{\mybig #1\over \mybig #2}}

\begin{document}
\thispagestyle{empty}

\baselineskip=12pt

{\small \noindent \mydate    \hfill OU-HET 494/2004}


\baselineskip=40pt plus 1pt minus 1pt

\vskip 3.cm

\begin{center}
{\Large \bf Dynamical Gauge-Higgs Unification}\\
{\Large \bf in the Electroweak Theory}\\


\vspace{3.cm}
\baselineskip=20pt plus 1pt minus 1pt

{\bf  Yutaka Hosotani\footnote{hosotani@het.phys.sci.osaka-u.ac.jp}, 
Shusaku Noda\footnote{noda@het.phys.sci.osaka-u.ac.jp} 
and Kazunori Takenaga\footnote{takenaga@het.phys.sci.osaka-u.ac.jp}}\\
\vspace{.3cm}
{\small \it Department of Physics, Osaka University,
Toyonaka, Osaka 560-0043, Japan}\\
\end{center}

\vskip 3.cm
\baselineskip=20pt plus 1pt minus 1pt

\begin{abstract}
$SU(2)_L$ doublet Higgs fields are unified with gauge fields
in the $U(3)_s \times U(3)_w$ model of Antoniadis, Benakli and 
Quir\'{o}s' on the orbifold $M^4 \times (T^2/Z_2)$. 
The effective potential for the Higgs fields (the Wilson 
line phases) is evaluated.   The electroweak symmetry is dynamically 
broken to $U(1)_{EM}$ by the Hosotani mechanism.  
There appear light Higgs particles. There is a phase transition as 
the moduli parameter of the complex structure of $T^2$ is varied.  
\end{abstract}

\newpage


\newpage

Gauge fields and Higgs scalar fields in four dimensions are unified 
in gauge theory in higher dimensions. In particular, gauge theory
defined in spacetime with  orbifold  extra dimensions has recently
attracted much attention in constructing  phenomenological models.

The idea of unifying Higgs scalar fields with gauge fields was 
first put forward  by Manton and Fairlie.\cite{Manton1} 
Manton considered $SU(3)$, $O(5)$ and $G_2$ 
gauge theory on $M^4 \times S^2$, supposing that field strengths
on $S^2$ are nonvanishing in such a way that gauge symmetry breaks down to
the electroweak $SU(2)_L \times U(1)_Y$.  Extra-dimensional components of 
gauge fields of the broken part of the symmetry are 
the Weinberg-Salam Higgs fields.  
Higher energy density resulting from nonvanishing field strengths 
on $S^2$, however, leads to the instability of the background 
configuration.  The stabilization of states with nonvanishing flux  
by quantum effects has been discussed.\cite{YH4}

The problem of the instability is more naturally solved by considering 
gauge theory on non-simply connected space.  It was shown \cite{YH1,YH2}
that quantum dynamics of Wilson line phases can induce gauge symmetry 
breaking.   In particular it was proposed to identify adjoint Higgs 
fields in grand unified theory (GUT) 
 with extra-dimensional  components of gauge fields, 
 which dynamically  induces the  symmetry breaking such as 
$SU(5) \go SU(3)_c \times SU(2)_L \times U(1)_Y$.

Significant progress along this line has been made recently by constructing
gauge theory on orbifolds.\cite{Pomarol1}-\cite{HNT}
  In gauge theory on orbifolds, 
boundary conditions imposed at fixed points on orbifolds incorporate 
a new way of gauge symmetry breaking.  With this orbifold symmetry breaking
some of light modes in the Kaluza-Klein tower expansion of 
 fields are eliminated from the spectrum at
low energies so that chiral fermions in four dimensions 
naturally emerge.  Furthermore, GUT on orbifolds can provide an elegant
solution to the triplet-doublet mass splitting problem of the Higgs 
fields \cite{Kawamura} and the gauge hierarchy problem.\cite{Lim2}
There is an attempt to unify  all of the gauge fields, Higgs fields and 
quarks and leptons as well.\cite{Nandi1}

In gauge theory on an orbifold,  boundary conditions given at the fixed points
of the orbifold play an important role.  This advantage, however, 
also implies indeterminacy in theory, namely the arbitrariness problem 
of boundary conditions.\cite{YH5}  It is desirable to show how a particular
set of boundary conditions is chosen naturally or dynamically.

It has been known that in gauge theory 
on non-simply connected space, different sets of boundary conditions 
can be physically  equivalent by the Hosotani mechanism.\cite{YH2}
It was shown in ref.\ \cite{HHHK} that there are  equivalence 
relations among different sets of boundary conditions  
in gauge theory on orbifolds as well.  In each equivalence class 
of boundary conditions, physics is independent of
boundary conditions imposed.  The physical symmetry is determined
by the dynamics of surviving Wilson line phases. Thus the arbitrariness
problem of boundary conditions is partly solved.

Dynamics of Wilson line phases are very important in the gauge-Higgs
unification in the electroweak theory.\cite{Lim1,HHHK}
In 2001, Antoniadis, Benakli and  Quir\'{o}s proposed an intriguing 
model of electroweak interactions.\cite{Antoniadis1}
$U(3)_s\times U(3)_w$ gauge theory is defined on 
$M^4 \times (T^2/Z_2)$.
In this model the weak gauge symmetry 
$U(3)_w \simeq SU(3)_w \times U(1)_w$ is broken, by the orbifold
boundary conditions, to $SU(2)_L \times U(1)_{w'} \times U(1)_w$.
The strong group $U(3)_s$ is decomposed as $SU(3)_c \times U(1)_s$.
Quarks and leptons are introduced such that among three $U(1)$ groups
only one combination, $U(1)_Y$, is free from anomalies, while
the gauge fields of the other two $U(1)$'s  become massive
by the Green-Schwarz mechanism.  Thus the 
surviving symmetry at the orbifold scale is 
$SU(3)_c \times SU(2)_L \times U(1)_Y$.

An amusing feature of this model is that a part of the 
extra-dimensional components of $SU(3)_w$ gauge fields
become $SU(2)_L$ doublet Higgs fields in four dimensions.
They are massless at the tree level.  They may acquire
nonvanishing vacuum expectation values at the one loop level,
thus breaking the electroweak symmetry to $U(1)_{EM}$.
At the same time they acquire nonvanishing masses.
These points were left unsettled in the original paper by
Antoniadis et al.
The purpose of this paper is to evaluate the effective potential
for the Higgs fields (the Wilson line phases) and examine
the resultant spectrum.  We find below that physics depends on
the matter content and also the moduli parameter of the complex
structure of $T^2$.  We will observe that light Higgs particles appear.

The model is defined on $M^4 \times (T^2/Z_2)$.  Let
$x^\mu$ $(\mu = 0, \cdots, 3)$ and $\vec y = (y^1, y^2)$ be
coordinates of $M^4$ and $T^2$, respectively.  Loop translations 
along the $y^a$ axis is given by $\vec y \go \vec y + \vec l_a$
where $\vec l_1 = (2\pi R_1, 0)$ and $\vec l_2 = (0 , 2\pi R_2)$.
The metric of $T^2$ is given by
\beqn
g_{ij}= \pmatrix{1 & \cos\theta \cr \cos\theta & 1} ~~,
\label{metric1}
\eeqn
where $\theta$ is the angle between the directions of the $y^1$ and $y^2$
axes.   The orbifold $T^2/Z_2$ is obtained by the 
$Z_2$ orbifolding, namely identifying  $\vec y $ with 
$- \vec y$.  The $Z_2$ orbifolding yields four fixed points
on $T^2/Z_2$; $\vec z_0 = \vec 0$, $\vec z_1 = \onehalf \vec l_1$, 
$\vec z_2 = \onehalf \vec l_2$, and 
$\vec z_3 = \onehalf (\vec l_1+ \vec l_2)$.

The Lagrangian density must be single-valued; 
${\cal L} (x, \vec y + \vec l_a) = {\cal L} (x, \vec y)$ $(a=1,2)$ and
${\cal L} (x, \vec z_j - \vec y) = {\cal L} (x, \vec z_j + \vec y)$ 
$(j=0, \cdots, 3)$.  However, this does not imply that 
fields are single-valued.  Instead, it is sufficient for gauge fields,
for instance,  to satisfy \cite{HNT}
\beqn
&&\hskip -1cm 
A_M(x, \vec y + \vec l_a) =
U_a A_M(x, \vec y ) \, U_a^\dagger  ~~,~~ (a=1,2) \cr
\noalign{\kern 10pt}
&&\hskip -1cm 
\pmatrix{A_\mu \cr A_{y^I} \cr} (x, \vec z_i - \vec y) =
P_i \pmatrix{A_\mu \cr - A_{y^I} \cr} (x, \vec z_i + \vec y) \, 
   P_i^\dagger  ~~,~~ (i = 0, \cdots, 3)
\label{bcg}
\eeqn
where $U_a^\dagger = U_a^{-1}$ and $P_i^\dagger = P_i= P_i^{-1}$.
The commutativity of two independent loop translations
demands $U_1 U_2 = U_2 U_1$.  Not all of $U_a$ and $P_i$ are 
independent; $U_a = P_0 P_a$ and 
$P_3 = P_2 P_0 P_1 = P_1 P_0 P_2$.

\def\myb{{\vphantom{\myfrac{1}{2}}}}
Let $g_s$ and $g$ be gauge coupling constants  for the 
groups $U(3)_s$ and $U(3)_w$, respectively.
The boundary conditions are given by
\beeq
P_0 = P_1 = P_2 = 
\Ithree  \otimes \pmatrix{-1\cr &-1 \cr &&+1\cr} ~~.
\label{BCmatrix1}
\eneq
Note that $U_1 = U_2 = \Ithree  \otimes \Ithree$, that is, gauge fields
are periodic on $T^2$. With the given boundary conditions 
$SU(3)_w$ symmetry breaks down to $SU(2)_L \times U(1)_{w'}$ at the classical
level.  There are zero modes of $A_{y^I}$, Wilson line phases, on $T^2$.
They are
\beeq
A_{y^I} ={1\over \sqrt{2} } \pmatrix{~ & \myb \Phi_I \cr
         ~~\Phi_I^\dagger ~ &  \cr} \quad (I=1,2)~.
\label{wilson1}
\eneq
$\Phi_1$ and $\Phi_2$ are $SU(2)_L$ doublets.  At the tree level 
the Lagrangian density  for the zero-modes of $\Phi_I$ is given by 
\beqn
&&\hskip -1cm
{\cal L}_{\rm tree}(\Phi_1, \Phi_2)
= g^{jk} (D^\mu \Phi_j)^\dagger (D_\mu \Phi_k) 
   - V_{\rm tree}(\Phi_1, \Phi_2) \cr
\noalign{\kern 10pt}
&&\hskip -1cm  
V_{\rm tree} = {g^2 \over 2 \sin^2 \theta} 
\Big\{ 
 \Phi_1^\dagger \Phi_1^{} \cdot \Phi_2^\dagger \Phi_2^{} 
+ \Phi_2^\dagger \Phi_1^{} \cdot \Phi_1^\dagger \Phi_2^{}
-(\Phi_2^\dagger \Phi_1^{})^2 - (\Phi_1^\dagger \Phi_2^{})^2 
\Big\} ~,
\label{tree1}
\eeqn
which has flat directions for $\Phi_1 = \gamma \Phi_2$ ($\gamma$: real),
corresponding to the vanishing field strength $F_{y^1 y^2} = 0$.
The potential $V_{\rm tree}$ does not have the custodial symmetry.

On $T^2/Z_2$ fermions satisfy
\beqn
\psi(x, \vec z_j - \vec y) = \eta_j \, T[P_j] \, (i\Gamma^4\Gamma^5)
\psi(x, \vec z_j + \vec y) ~~~~~\quad (j=0,1, 2,3) ~~
\label{bcf}
\eeqn
Here  
$T[P_j]$ stands for an appropriate representation matrix under 
the gauge group associated with $P_j$. If $\psi$ belongs to the 
fundamental  representation, $T[P_j] \, \psi = P_j \psi$. 
$\eta_j = \pm 1$ and $\eta_3 = \eta_0 \eta_1 \eta_2$.  
$\{ \Gamma^a ; (a=0 \sim 5) \}$  are six dimensional $8\times 8$ 
Dirac's  matrices. Four- and six-dimensional chiral operators are 
given by $\Gamma^{4c} = i \Gamma^0 \cdots \Gamma^3$ and
$\Gamma^{6c} = i \Gamma^{4c}\Gamma^4 \Gamma^5$, respectively. 

Quarks and leptons are introduced in such a way that only one 
combination of three
$U(1)$'s remains free from anomaly.\cite{Antoniadis1}  
They come in as six-dimensional (6D) Weyl fermions.  Three families of 
fermions are introduced as
\beqn
&&\hskip -1cm 
L_{1,2,3}=({\bf 1,3})^{+} ~~,~~
 D^{c}_{1,2,3}=({\bf \bar{3} ,1})^{+} ~~,  \cr
&&\hskip -1cm 
Q_1=({\bf 3,\bar{3}})^{+} ~~,~~   
Q_2=({\bf 3,\bar{3}})^{-} ~~,~~
Q_3=({\bf \bar{3},3})^{-} 
\label{assignment1}
\eeqn
where $({\bf n_s, n_w})^{\epsilon}$ stands for a  fermion 
with 6D chirality $\epsilon =\pm$ in the representations  
${\bf n_s}$ and ${\bf n_w}$ of $U(3)_s$  and $U(3)_w$, respectively. 
Each 6D Weyl fermion is decomposed into 4D left-handed (L) and 
right-handed (R) fermions.  $L_1$ in (\ref{assignment1}),
for instance, consists of $L_{1L} = (\nu_L, e_L, \tilde e_L)$
and $L_{1R} = (\tilde \nu_R, \tilde e_R, e_R)$.  Among them  
$\nu_L$, $e_L$ and $e_R$ have zero modes, whereas 
$\tilde \nu_R$, $\tilde e_L$ and $\tilde e_R$ do not.
Let ${\cal Q}_c$, ${\cal Q}_w$, and ${\cal Q}_{w'}$ be appropriately
normalized charges of $U(1)_c$, $U(1)_w$, and $U(1)_{w'}$.
It has been shown in ref.\ \cite{Antoniadis1} that
with the assignment (\ref{assignment1}), the charge 
${\cal Q}_Y =  -\frac{1}{3} {\cal Q}_c 
-\frac{2}{3} {\cal Q}_w + \frac{1}{6} {\cal Q}_{w^{\prime}}$ is
anomaly free.
${\cal Q}_Y$ is the weak hypercharge.  The resultant theory at low
energies is exactly  the standard Weinberg-Salam theory of
massless quarks and leptons with two Higgs 
doublets.  The weak hypercharge coupling $g_Y$ is
given by $g^{-2}_Y = 3 g^{-2} + \frac{2}{3} g_s^{-2}$. Should the 
electroweak symmetry breaking take place, 
the Weinberg angle  is given by
\beeq
\sin^2 \theta _w = {1 \over ~ 4+  \myfrac{2g^2}{3g^2_s} ~}~~.
\label{Wangle}
\eneq
which is close to the observed value.

The main question is if the electroweak symmetry breaking takes place
at the quantum level through the Hosotani mechanism.  The effective
potential $V_\eff$ for $\Phi_I$  becomes nontrivial even in the flat
directions of the  potential $V_{\rm tree}$ in  (\ref{tree1}). 
The minimum of $V_\eff$ can be  
at nontrivial values of $\Phi_I$,  the symmetry breaking being induced 
and  the Higgs fields acquiring finite masses.  
 
It is sufficient to evaluate the effective potential for a configuration
\beeq
\sqrt{2} g R_1 \Phi_1 = \pmatrix{0 \cr a\cr} ~~,~~
\sqrt{2} g R_2 \Phi_2 = \pmatrix{0 \cr b\cr} ~~,
\label{Wilson2}
\eneq
where $a$ and $b$ are phase variables with a period 2. 
Depending on the location of the global minimum of $V_\eff(a,b)$,
the physical symmetry varies.  To pin down the physical symmetry, 
it is most convenient to move to a new gauge, in which 
$\la A_{y^I}' \ra = 0$, by a gauge transformation
\beeq
\Omega (\vec y ~; a,b) = \exp \Bigg\{ i \bigg( 
  \myfrac{a y^1}{2 R_1} + \myfrac{b y^2}{2 R_2}
   \bigg) \lambda^6 \Bigg\} ~~.
\label{largeGT3}
\eneq
Then new parity matrices in (\ref{bcg})  become
\begin{equation}
P_0^{\prime}=\pmatrix{-1 & & \cr &-\tau_3},
\quad 
P_1^{\prime}=\pmatrix{ -1 &  \cr & -{\rm e}^{i\pi a\tau^1}\tau^3},\quad
P_2^{\prime}=\pmatrix{ -1 &  \cr & -{\rm e}^{i\pi b\tau^1}\tau^3} ~.
\label{newBC}
\end{equation}  
As shown in \cite{HNT}, generators commuting with the 
new $P_i^{\prime}$ $(i=0, 1, 2)$ span the  algebra 
of the physical symmetry.  The physical symmetry is given by
\beeq
\cases{SU(2)_L \times U(1)_Y
&for $(a,b) = (0,0)$,\cr
\noalign{\kern 5pt}
U(1)_{EM} \times U(1)_Z
&for $(a,b) = (0,1), (1,0), (1,1)$,\cr
\noalign{\kern 5pt}
U(1)_{EM} &otherwise.\cr}
\label{symmetry1}
\eneq
For generic values of $(a,b)$, electroweak symmetry breaking
takes place and the Weinberg angle is given by (\ref{Wangle}). 

The evaluation of the effective potential is reduced to finding
contributions from a $Z_2$-doublet $\phi^t = (\phi_1, \phi_2)$.\cite{HHHK} 
When $\phi$ satisfies boundary conditions
 \beeq
 \phi (x, -\vec y) = \tau_3 \phi(x, \vec y) ~~,~~
 \phi (x, \vec y + \vec l_a) = e^{2 \pi i \gamma_a \tau_2} \phi(x, \vec y) ~~,
 \label{doublet1}
 \eneq
 its mode expansion is given by
 \beqn
 &&\hskip -1cm
 \phi(x, \vec y) = {1\over \sqrt{2\pi^2 R_1 R_2 \sin\theta}}
 \sum_{n,m=-\infty}^\infty 
 \phi_{nm}(x) \pmatrix{ \cos c_{nm}(\vec y) \cr \sin c_{nm}(\vec y) \cr} \cr
 \noalign{\kern 5pt}
 &&\hskip -.0cm
 c_{nm}(\vec y) = \myfrac{(n+\gamma_1) y_1}{R_1} 
                + \myfrac{(m+\gamma_2) y_2}{R_2} ~~.
 \label{doublet2}
 \eeqn
 The Lagrangian density for $\phi$, including the interaction with
 Wilson line phases $\alpha_j$, is given by
\beeq
{\cal L}_1 = \myfrac{1}{2}  g^{jk} D_j \phi_a  D_k \phi_a  ~~,~~
D_j \phi_a = \dd_j \phi_a - \myfrac{\alpha_j}{R_j} \ep_{ab} \phi_b ~~.
\label{doublet3}
\eneq
Inserting (\ref{doublet2}) into $\int d^2 y \sqrt{ g} \, {\cal L}_1$, 
one obtains the spectrum of $\phi_{nm}(x)$ fields.  From the spectrum  
the contribution of
$[ \phi_1, \phi_2 ; (\gamma_1, \gamma_2), (\alpha_1, \alpha_2)]$ to the
1-loop effective potential is found to be 
$I[\alpha_1 + \gamma_1, \alpha_2 + \gamma_2 ; \cos\theta]$ 
where \cite{LeeHo, Antoniadis1}
\beqn
&&\hskip -1cm
I[\alpha, \beta; \cos \theta] 
= - {1\over 16\pi^9} 
\Bigg\{ \sum_{n=1}^\infty \myfrac{\cos 2 \pi n \alpha}{n^6 R_1^6}
       + \sum_{m=1}^\infty \myfrac{\cos 2 \pi m \beta}{m^6 R_2^6} \cr
\noalign{\kern 10pt}
&&\hskip 1.cm
+ \sum_{n=1}^\infty  \sum_{m=1}^\infty  
\myfrac{\cos 2\pi (n\alpha + m \beta)} 
   {(n^2 R_1^2 + m^2 R_2^2 + 2  nmR_1 R_2 \cos \theta)^3}    \cr  
\noalign{\kern 10pt}
&&\hskip 1.cm
+ \sum_{n=1}^\infty  \sum_{m=1}^\infty
\myfrac{\cos 2\pi (n\alpha - m \beta)}
   {(n^2 R_1^2 + m^2 R_2^2 - 2  nmR_1 R_2 \cos \theta)^3} \Bigg\}  ~.
\label{Ifunction1}
\eeqn     
One unit of $I$ represents contributions to the effective
potential from two physical degrees of freedom.
Note that $I[-\alpha, \beta; \cos \theta] = I[\alpha, -\beta; \cos \theta]
= I[\alpha, \beta; -\cos \theta]$. 

Contributions from $SU(3)_w$ gauge fields  to
the effective potential $V_\eff (a,b)$ in the background field gauge
are given, for each degree of freedom, by
$V_\eff^{\rm g + gh}=-  {i\over 2}{\rm tr~ln} \, D_L D^L(A_{y^I} )$,
as both the Ricci tensors and background gauge field strengths
vanish.\cite{YH2}  For ghost fields the sign is reversed.

For each spacetime component of gauge fields in the 
tetrad  frame we write $B = \sum_{a=1}^8 \onehalf B_a \lambda_a$. 
The four-dimensional components of gauge fields and ghost fields
decompose into the following $Z_2$-doublets 
$[\phi_1,\phi_2; (\gamma_1.\gamma_2),(\alpha_1.\alpha_2)]$;
\beqn
&&\hskip -1cm 
[- \onehalf(\sqrt{3} B_3 + B_8), B_6; (0,0), (0,0)] \cr
&&\hskip -1cm 
[ B_1 , B_5; (0,0), (\onehalf a, \onehalf b)] \cr
&&\hskip -1cm 
[ B_2 , B_4; (0,0), (-\onehalf a, -\onehalf b)] \cr
&&\hskip -1cm 
[- \onehalf( B_3 - \sqrt{3} B_8), B_7; (0,0), (a,b)] ~.
\label{effV1}
\eeqn
Similarly, the extra-dimensional components of gauge fields decompose into
\beqn
&&\hskip -1cm 
[B_6, - \onehalf(\sqrt{3} B_3 + B_8); (0,0), (0,0)] \cr
&&\hskip -1cm 
[B_5, B_1; (0,0), (-\onehalf a, -\onehalf b)] \cr
&&\hskip -1cm 
[ B_4, B_2 ; (0,0), (\onehalf a, \onehalf b)] \cr
&&\hskip -1cm 
[B_7, - \onehalf( B_3 - \sqrt{3} B_8); (0,0), (-a,-b)] ~.
\label{effV2}
\eeqn
Summing up all these contributions, one finds
\beeq
V_\eff(a,b)^{\rm g+gh} = 4 I(0,0) + 8 I(\onehalf a, \onehalf b) 
 + 4 I(a,b) ~~. 
\label{Veff1}
\eneq
Here $I(a,b) = I(a,b; \cos\theta)$.

To find contributions from fermions, one notes that the 
extra-dimensional part of the Dirac operator is given by
$\bar D = i \Gamma^a {e_a}^j D_j(A_{y^I})$ ($a=4,5$, $j= y^j$)
where the tetrad satisfies
$\delta^{ab} {e_a}_j {e_b}_k = g_{jk}$
and $D_j (A_{y^I})$ is a covariant derivative.  As $T^2$ is flat,
spin connections vanish.  To evaluate the effective potential 
$V_\eff(a,b)$ at one loop, it is sufficient to know the spectrum 
of $\bar D$.  As $i \Gamma^4 \Gamma^5 \psi$ satisfies the 
same boundary condition (\ref{bcf}) as $\psi$ and 
$i \Gamma^4 \Gamma^5 \bar D = - \bar D   i \Gamma^4 \Gamma^5$,
eigenvalues $\lambda(a,b)$ of $\bar D \psi = \lambda \psi$
always appear in a pair $(\lambda, -\lambda)$.  The exception is
for modes with $\lambda=0$ which give irrelevant constant contributions 
to  $V_\eff(a,b)$.  Hence contributions from fermions
are summarized as 
$+ \onehalf i \tr (D_4 + \bar D) = i \tr (D_4^2 + \bar D^2)$.

As the tetrads are constant and the background $F_{y^1 y^2}=0$,
$\bar D^2 = -g^{jk} D_j D_k$.  As in the case of bosons, nontrivial
contributions arise from $Z_2$ doublets.  From $L_1$, for instance, one has
\beqn
&&\hskip -1cm
[ e_L, i \tilde e_L ; (0,0), (\onehalf a, \onehalf b)] \cr
&&\hskip -1cm
[ e_R, -i \tilde e_R ; (0,0), (-\onehalf a, -\onehalf b)] ~~.
\label{effV3}
\eeqn
$\nu_L$ and $\tilde \nu_R$ have no coupling to $a$ and $b$.
Similar results are obtained for $L_j$ and $Q_j$.  $D_j^c$ does
not couple to $a$ and $b$.  To summarize, three families of fermions
give
\beeq
V_\eff(a,b)^{\rm f} = -3 \, \Big\{
14 I(0,0) + 16 I(\onehalf a, \onehalf b) \Big\} ~~.
\label{Veff2}
\eneq
Adding (\ref{Veff1}) and (\ref{Veff2}), one finds
\beeq
V_\eff(a,b )^{\rm total} = 
 -40 I(\onehalf a, \onehalf b )  
 + 4 I(a,b ) 
\label{Veff3}
\eneq
up to a constant.

\begin{figure}[tbh]
\centering
\leavevmode
\includegraphics[width=7.cm]{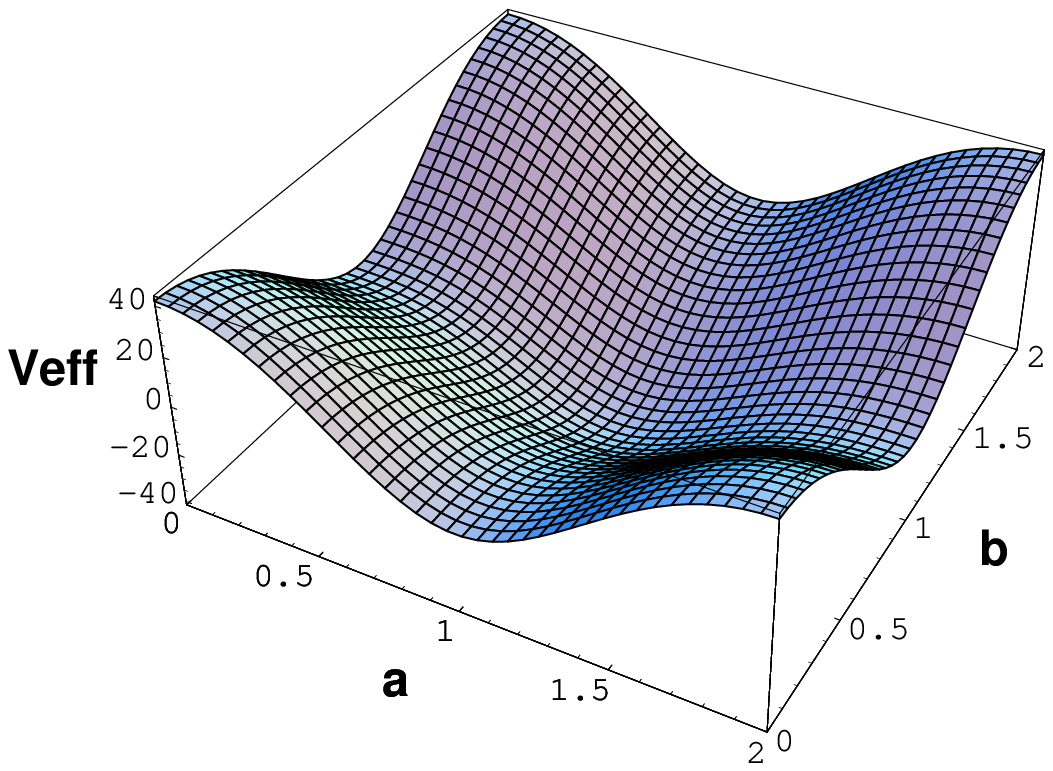}
\hskip 0.5cm
\includegraphics[width=7.cm]{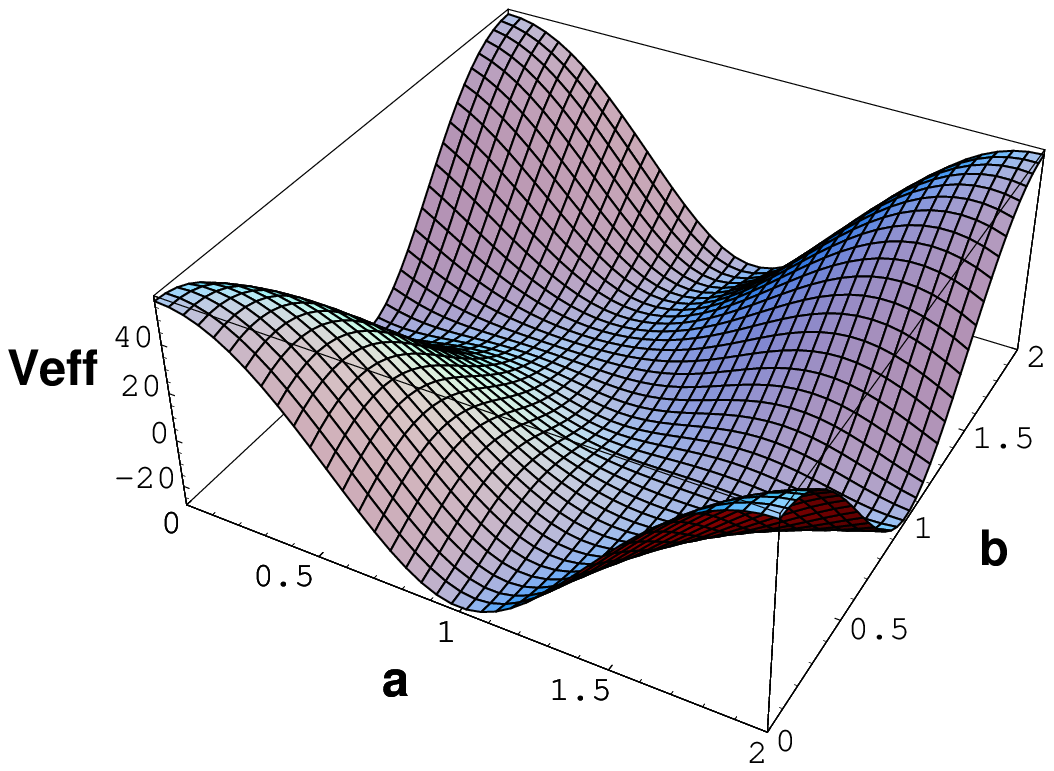}\\
{(a) \hskip 7cm (b)}
\caption{$V_\eff(a,b)$  in (\ref{Veff3}) for  $R_1=R_2$.
(a) $\cos\theta =0$.  (b) $\cos\theta =0.5$.}
\label{fig-V0}
\end{figure}

Given $R_1$, $R_2$ and $\cos \theta$, the absolute minimum of
$V_\eff(a,b)$ is easily found.  First note that in the pure
gauge theory $V_\eff(a,b)^{\rm g+gh}$ in (\ref{Veff1}) 
has the minimum at $(a,b)=(0,0)$, i.e.\ $SU(2)_L \times U(1)_Y$ 
symmetry remains unbroken.  In the presence of fermions the 
symmetry is partly broken.  For $\cos\theta =0$, the minimum of 
$V_\eff(a,b)^{\rm total}$ in (\ref{Veff3}) is located at 
$(a,b) = (1,1)$. See fig.\ \ref{fig-V0}(a).
The symmetry breaks down to $U(1)_{EM} \times U(1)_Z$.  
$Z$ bosons remain massless, which is not what is sought for.

Before presenting models with the electroweak symmetry breaking,
we would like to comment that the phase structure critically depends on 
the value of $\cos \theta$.  For $\cos\theta < 0.5$,
the absolute minimum is located at $(a,b)=(1,1)$.  At 
$\cos\theta = 0.5$ with $R_1=R_2$, there appear three degenerate
minima at $(a,b) = (1,1), (0,1), (1,0)$.  See fig.\ \ref{fig-V0}(b).
Notice that the barrier height $V_{\rm B}$ separating
the three minima is very small compared with the potential height.
For $\cos\theta > 0.5$, the absolute minima are given by
$(a,b) = (0,1), (1,0)$.  
Although the physical symmetry in the model leading to $V_\eff$ in (\ref{Veff3}) is 
$U(1)_{EM} \times U(1)_Z$ for all values of $\cos\theta$, 
 the spectrum changes at $\cos\theta=0.5$.  There is a 
first-order phase transition there.
 \ignore{
 Even in the case $R_1=R_2\equiv R$, there appear two distinct
energy scales.  One scale is set by 
$(\hbox{volume of } T^2)^{-1/2} = (R\sin\theta)^{-1}$, which 
characterizes the Kaluza-Klein towers including the $W$ boson mass.
The other is given by the curvature at the minimum of $V_\eff$
which gives the mass of the light Higgs field.
}

Models with the electroweak symmetry breaking are obtained 
by adding heavy fermions.  For each quark/lepton multiplet in 
(\ref{assignment1}), which has $(\eta_0 \eta_1, \eta_0 \eta_2) = (1,1)$
in (\ref{bcf}),  we introduce three parity partners with
$(\eta_0 \eta_1, \eta_0 \eta_2) = (-1,1), (1,-1), (-1,-1)$. Further we
add fermions in the adjoint representation with 
$(\eta_0 \eta_1, \eta_0 \eta_2) = (-1,1)$. The total effective potential is,
up to a constant, 
\beqn
&&\hskip -1cm
V_\eff(a,b)^{\rm total} = 
8 I(\onehalf a, \onehalf b)  + 4 I(a,b) 
- N_{Ad} \Big\{ 8 I(\onehalf a + \onehalf, \onehalf b)  
   + 4 I(a + \onehalf,b) \Big\} \cr
\noalign{\kern 10pt}
&&\hskip 0.cm
 - 16 N_F \Big\{ I(\onehalf a, \onehalf b )  
 + I(\onehalf a + \onehalf, \onehalf b)
 + I(\onehalf a , \onehalf b + \onehalf)
 + I(\onehalf a + \onehalf, \onehalf b + \onehalf) \Big\} ~.
\label{Veff4}
\eeqn
Here $N_{Ad}$ and $N_F$ are the numbers of Weyl fermions in the adjoint
representation and of generation of quarks and leptons, respectively.
Fermions with $(\eta_0 \eta_1, \eta_0 \eta_2) \not= (1,1)$ do not have
zero modes.  For $N_F=3$ the spectrum at low energies is the same as 
in the minimal model leading to (\ref{Veff3}).

\begin{figure}[tbh]
\centering
\leavevmode
\includegraphics[width=7.cm]{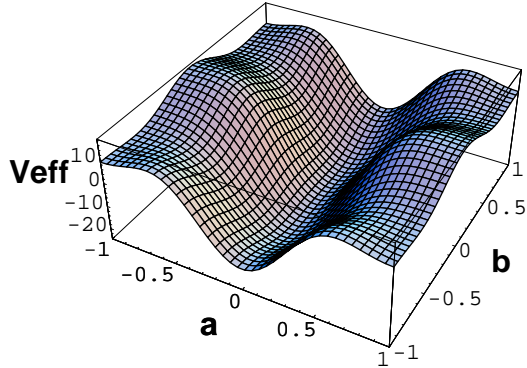}
\caption{$V_\eff(a,b)$  in (\ref{Veff4})  with $N_{Ad}=1$, 
$N_F=3$, $\cos\theta =0$ and $R_1=R_2$.  The minimum is located at 
 $(a,b)=(0, \pm 0.269)$. Dynamical electroweak symmetry breaking takes place.}
\label{fig-Ad=1}
\end{figure}

An interesting model is obtained for $N_{Ad}=1$ and $N_F=3$.  
$V_\eff$ with $N_{Ad}=1$, $N_F=3$, $\cos\theta =0$ and $R_1=R_2$
is displayed in fig.\ \ref{fig-Ad=1}.  The global minima are
located at $(a,b)=(0, \pm 0.269)$.  The $SU(2)_L \times U(1)_Y$ symmetry
breaks down to $U(1)_{EM}$.  At $\cos \theta = 0.1$ the global minima 
move to $(a,b)=(\pm 0.013, \pm 0.224)$.  There is a critical value for
$\cos\theta$.  At $\cos \theta  = 0.133 \equiv \cos \theta_c$,  
the minima at 
$(a,b)=(\pm 0.0135, \pm 0.158)$ become degenerate with the minimum
at $(a,b)=(0,0)$.  For $\cos \theta > \cos \theta_c$, the point $(a,b)=(0,0)$ is 
the  global minimum and the symmetry remains unbroken.  
There is a first-order phase transition at $\cos \theta_c$.
We note that dynamical electroweak symmetry breaking takes place for
$N_{Ad} =0$ and $N_F \ge 7$.  For instance,
the global minima of $V_\eff$ for $N_{Ad}=0$, $N_F=9$, $\cos\theta=0$, 
and $R_1=R_2$ are located at $a= \pm b$,  $a= \pm 0.320$.

Let us examine the spectrum of gauge bosons and Higgs particles in the 
model $N_{Ad}=1$, $N_F=3$.  The mass of $W$ bosons is given by
\beeq
m_W^2 =
\frac{1}{4 \sin^2\theta} 
\left( \frac{a_0^2}{R_1^2} +\frac{b_0^2}{R_2^2} 
- \frac{2a_0 b_0 \cos\theta }{R_1 R_2} \right)  ~. 
\label{spectrum1}
\eneq
When $R_1=R_2=R$, 
$m_W= 0.135 R^{-1}$ and $0.112  R^{-1}$ for 
$\cos\theta =0$ and  $0.1$, respectively.
Here $(a_0, b_0)$ denotes the location of the global minimum of $V_\eff$.

The mass of $Z$ bosons is subtle. For $\cos\theta < \cos \theta_c$,
only photons ($A_\mu^{EM}$) remain massless.  Let
$A_\mu^{Y}=A_\mu^{Y_1}$, $A_\mu^{Y_2}$ and $A_\mu^{Y_3}$ be gauge 
fields associated  with the weak hypercharge $Y=Y_1$ and the other 
two $U(1)$ charges $Y_2$ and $Y_3$.
These three gauge fields are related to $A_\mu^{w (8)}$ and the two  
gauge fields associated with $U(1)_s$ and $U(1)_w$ by an orthogonal 
transformation. In particular
$A_\mu^{w (8)} = \sum_{j=1}^3 \Omega_{1j} A_\mu^{Y_j} $ where
$\Omega \in SO(3)$ and $\sqrt{3} \, \Omega_{11} = \tan \theta_W$.
The (mass)$^2$ matrix in the basis $(A_\mu^{w (3)}, A_\mu^{Y_j})$ is 
given, for $\cos\theta=0$,  by
\beqn
&&\hskip -1cm
K = K^{(0)} + K^{(1)}\cr
\noalign{\kern 10pt}
&&\hskip -.5cm
= \pmatrix{0 \cr & 0\cr && m_2^2 \cr &&& m_3^2}
+ m_W^2 \pmatrix{ 1 & - \sqrt{3} \Omega_{1k} \cr \cr
         - \sqrt{3} \Omega_{1j} & 3 \Omega_{1j} \Omega_{1k} } ~.
\label{rho1}
\eeqn
Here $m_2$ and $m_3$ are the masses of $A_\mu^{Y_2}$ and $A_\mu^{Y_3}$
acquired through the Green-Schwarz mechanism.  The second term
$K^{(1)}$ arises from the $\onehalf \Tr (F^{w}_{\mu y_j})^2$ term
 with nonvanishing $\la A^{w}_{y_j} \ra$. We suppose that 
 $m_2^2, m_3^2 \gg m_W^2$. 
 
One of the eigenstates of $K$ is $A_\mu^{EM}$ which has an eigenvector 
$\vec v_{EM}{}^t = (\cos\theta_W, - \sin\theta_W, 0, 0)$ with a vanishing
eigenvalue.  The Weinberg angle is given by (\ref{Wangle}). 
The eigenvalue and eigenvector for a $Z$ boson are found
in a power series in $m_W^2/m_j^2$.  One finds
\beeq
m_Z^2 = \myfrac{m_W^2}{\cos \theta_W^2} \,
\Bigg\{ 1 - \bigg(\Omega_{12}^2 \myfrac{m_W^2}{m_2^2} +
       \Omega_{13}^2 \myfrac{m_W^2}{m_3^2} \bigg) \Bigg\} ~.
\label{rho2}
\eneq
Due to the mixing with $A_\mu^{Y_2}$ and $A_\mu^{Y_3}$, the 
$\rho$ parameter ($= m^2_{W}/(m^2_{Z} \cos^2 \theta_W)$) 
becomes slightly bigger than 1 at the tree level.  The correction
remains small if the masses generated by the Green-Schwarz mechanism are
much larger than $m_W$. 

A prominent feature of the model is observed in the spectrum of the Higgs 
particles.  In the six-dimensional model there are two Higgs doublets,
$\Phi_1$ and $\Phi_2$.  
With parametrization
\beeq
\Phi_j =  \pmatrix{ H_j \cr 2^{-1/2} (v_j + \phi_j + i \chi_j) }
\eneq
where $(v_1, v_2) = (a_0/gR_1, b_0/gR_2)$, 
the bilinear terms in $V_{\rm tree}$ are given by
\beqn
&&\hskip -1cm 
V_{\rm tree}=
{g^2\over 4 {\sin^2\theta}}(H_1^\dagger, H_2^\dagger)
\pmatrix{v_2^2 & -v_1v_2 \cr -v_1v_2 & v_1^2 }
\pmatrix{H_1 \cr H_2}  \cr
\noalign{\kern 10pt}
&&\hskip 0.2cm 
+ {g^2\over  2{\sin^2\theta}}
(\chi_1, \chi_2)
\pmatrix{v_2^2 & -v_1v_2 \cr -v_1v_2 & v_1^2 }
\pmatrix{\chi_1 \cr \chi_2}   ~~.
\label{mass2}
\eeqn
Among charged Higgs fields $(H_1, H_2)$, there are two massless modes and 
two massive modes with  a mass $ g(v_1^2 + v_2^2)^{1/2} /2 \sin\theta$.  
Neutral CP-odd Higgs fields $(\chi_1, \chi_2)$ decompose into one massless 
mode and one massive mode with a mass $ g(v_1^2 + v_2^2)^{1/2} /\sin\theta$.  
The three massless modes are absorbed by $W$ and $Z$.  For $\cos\theta = 0$
physical charged Higgs particles and neutral CP-odd Higgs particle have
masses $m_W$ and $2 m_W$, respectively.

Neutral CP-even Higgs particles $(\phi_1, \phi_2)$ are massless at the
tree level,  but do acquire finite masses from $V_\eff^{\rm 1-loop}$.
Noting that $a_j = a_{j0} + gR_j \phi_j$, where $(a_1,a_2) \equiv (a,b)$,  in
$V_\eff(a_1,a_2)$, the effective Lagrangian density for 
the zero modes of $\phi_j$ is
\beqn
&&\hskip -1cm
{\cal L}_\eff = {1\over 2} g^{jk} \dd_\mu \phi_j \dd^\mu \phi_k
 - {1\over 2} K^{jk} \phi_j \phi_k ~~, \cr
 \noalign{\kern 5pt}
&&\hskip -1cm 
K^{jk} = g^2 R_j R_k  
\left. \myfrac{\dd^2 V_\eff}{\dd a_j \dd a_k} \right|_{\rm min} ~~. 
\label{mass3}
\eeqn
Hence the two eigenvalues of (mass)$^2$ are
$\onehalf \big( A \pm \sqrt{A^2 - 4 B\sin^2 \theta} \big)$
where $A = g_{jk} K^{jk}$ and $B = \det K$.
In the case $N_{Ad}=1$, $N_F=3$, and $R_1=R_2=R$ in (\ref{Veff4}),
one of the CP-even Higgs particles is much heavier than the other.
Let us denote the four-dimensional gauge coupling by 
$g_4^2 = g^2/(2\pi^2 R^2 \sin\theta)$.  For $\cos\theta=0$, the masses are
given by $(0.871, 3.26) \times \sqrt{\alpha_w} \, m_W$ where 
$\alpha_w = g_4^2/4\pi$.  For $\cos\theta=0.1$, they are
$(0.799, 4.01) \times \sqrt{\alpha_w} \, m_W$.
In the case $N_{Ad}=0$, $N_F=9$ and
$\cos\theta =0$ in (\ref{Veff4}),  the Higgs masses are given by 
$(1.039, 1.174) \times \sqrt{\alpha_w} \, m_W$.
In the current scheme the mass of the lightest Higgs particle comes 
out too low.

In the dynamical gauge-Higgs unification scheme in six dimensions,
O($m_W$) masses  of charged and CP-odd neutral Higgs fields are
generated from  the $\Tr F_{y^1 y^2}^2$ term, or $V_{\rm tree}(\Phi)$ 
in (\ref{tree1}).
Along the flat directions in $V_{\rm tree}$, there appear light 
CP-even neutral Higgs fields.  Their masses squared are generated at the 
one loop level and therefore are suppressed by a factor $\alpha_w$.  

As for the masses of quarks and leptons, the current model yields
a mass spectrum which is independent of the generations, 
and therefore is not realistic.  
As pointed out in ref.\ \cite{YH3},  each fermion multiplet can 
acquire a mass from $Z_2$-twists in the boundary conditions 
on $T^2$. By combining this with the VEV of $\Phi_j$, 
it may be possible to produce hierarchy in the mass spectrum.

In this paper we have examined the $U(3) \times U(3)$ model of Antoniadis 
et al.\ to find that the electroweak symmetry breaking dynamically 
takes place through the Hosotani mechanism, 
provided additional heavy fermions are added. Higgs fields are 
unified with gauge fields.  There appear both
neutral and charged Higgs particles at the weak scale.  The Weinberg angle 
comes out about the observed value.

Although it is very encouraging that dynamical electroweak symmetry 
breaking is implemented in the scheme of the gauge-Higgs unification,
there remain several issues to be addressed.  First, the Kaluza-Klein 
mass scale ($1/R$) turns out to be about $10 m_W$, which is too small.
This is a general feature of the  models constructed on flat space.
The effective potential is minimized at O(1) values of the Wilson line
phases (Higgs fields).  The Higgsless model on the 
Randall-Sundrum background\cite{Csaki1} and the Hosotani mechanism on 
the warped spacetime\cite{Oda1} may have a hint for the resolution
of this problem.
Secondly,  the fermion mass spectrum in the model does not distinguish
one generation from the others.  One of the roles of the 
Higgs doublets in the Weinberg-Salam theory is to give fermions
masses.  As the Higgs fields become a part of gauge fields in the 
gauge-Higgs unification scenario,  additional sources for fermion
masses need to be introduced. 
Thirdly the dynamical 
gauge-Higgs unification scheme generically yields light Higgs
particles which may conflict with experimental data.
Fourthly it is desirable to extend the model to supersymmetric cases
where the Scherk-Schwarz SUSY breaking can induce gauge symmetry 
breaking.\cite{Takenaga1}  Further, in six dimensions one can introduce
bare fermion mass terms as well whose quantum effect on gauge symmetry 
breaking is of great interest.\cite{Takenaga2}
We hope to come back to these issues in separate 
publications.

\vskip 1cm

\leftline{\bf Acknowledgments}
This work was supported in part by  Scientific Grants from the Ministry of 
Education and Science, Grant No.\ 13135215, Grant No.\ 13640284, and
Grant No.\ 15340078 (Y.H), and by the 21st Century COE Program at Osaka 
University (K.T.).



\vskip 1.cm

\def\jnl#1#2#3#4{{#1}{\bf #2} (#4) #3}

\def\Zphys{{\em Z.\ Phys.} }
\def\jssc{{\em J.\ Solid State Chem.\ }}
\def\jpsJ{{\em J.\ Phys.\ Soc.\ Japan }}
\def\ptps{{\em Prog.\ Theoret.\ Phys.\ Suppl.\ }}
\def\PTP{{\em Prog.\ Theoret.\ Phys.\  }}

\def\JMP{{\em J. Math.\ Phys.} }
\def\NPB{{\em Nucl.\ Phys.} B}
\def\NP{{\em Nucl.\ Phys.} }
\def\PLB{{\em Phys.\ Lett.} B}
\def\PL{{\em Phys.\ Lett.} }
\def\PRL{\em Phys.\ Rev.\ Lett. }
\def\PRB{{\em Phys.\ Rev.} B}
\def\PRD{{\em Phys.\ Rev.} D}
\def\PRe{{\em Phys.\ Rep.} }
\def\AP{{\em Ann.\ Phys.\ (N.Y.)} }
\def\RMP{{\em Rev.\ Mod.\ Phys.} }
\def\ZPC{{\em Z.\ Phys.} C}
\def\SCI{\em Science}
\def\CMP{\em Comm.\ Math.\ Phys. }
\def\MPLA{{\em Mod.\ Phys.\ Lett.} A}
\def\IJMPA{{\em Int.\ J.\ Mod.\ Phys.} A}
\def\IJMPB{{\em Int.\ J.\ Mod.\ Phys.} B}
\def\EPJC{{\em Eur.\ Phys.\ J.} C}
\def\PR{{\em Phys.\ Rev.} }
\def\JHEP{{\em JHEP} }
\def\cmp{{\em Com.\ Math.\ Phys.}}
\def\JPA{{\em J.\  Phys.} A}
\def\JPG{{\em J.\  Phys.} G}
\def\NJP{{\em New.\ J.\  Phys.} }
\def\CQG{\em Class.\ Quant.\ Grav. }
\def\ATMP{{\em Adv.\ Theoret.\ Math.\ Phys.} }
\def\ibid{{\em ibid.} }

\renewenvironment{thebibliography}[1]
         {\begin{list}{[$\,$\arabic{enumi}$\,$]}  
         {\usecounter{enumi}\setlength{\parsep}{0pt}
          \setlength{\itemsep}{0pt}  \renewcommand{\baselinestretch}{1.2}
          \settowidth
         {\labelwidth}{#1 ~ ~}\sloppy}}{\end{list}}

\def\reftitle#1{}                


\begin{thebibliography}{99}
\small
\baselineskip=14pt


\leftline{\bf References}



\bibitem{Manton1}
N.\ Manton, \jnl{\NPB}{158}{141}{1979};
\reftitle{A New Six-Dimensional Approach To The Weinberg-Salam Model}

D.B.\ Fairlie, \jnl{\PLB}{82}{97}{1979};
\reftitle{Higgs' Fields And The Determination Of The Weinberg Angle}
\jnl{\JPG}{5}{L55}{1979};
\reftitle{Two Consistent Calculations Of The Weinberg Angle}

P.\ Forgacs and N.\ Manton, \jnl{\CMP}{72}{15}{1980}.
\reftitle{Space-Time Symmetries In Gauge Theories}

\bibitem{YH4}
Y.\ Hosotani,  \jnl{\PLB}{129}{193}{1984}; 
\reftitle{Dynamical Gauge Symmetry Breaking As The Casimir Effect}
\jnl{\PRD}{29}{731}{1984}.
\reftitle{Dynamical Gauge Symmetry Breaking And Left-Right Asymmetry In Higher Dimensional Theories}

\bibitem{YH1}
Y.\ Hosotani, \jnl{\PLB}{126}{309}{1983}.
\reftitle{Dynamical Mass Generation By Compact Extra Dimensions}

\bibitem{YH2}
Y.\ Hosotani, \jnl{\AP}{190}{233}{1989}.
\reftitle{Dynamics Of Nonintegrable Phases And Gauge Symmetry Breaking}

\bibitem{Pomarol1}
A.\ Pomarol and M.\ Quiros, \jnl{\PLB}{438}{255}{1998}.
\reftitle{The Standard Model from extra dimensions}

\bibitem{Lim2}
H.\ Hatanaka, T.\ Inami and C.S.\ Lim, 
\jnl{\MPLA}{13}{2601}{1998};
\reftitle{The Gauge Hierarchy Problem and Higher Dimensional Gauge Theories}

K.\ Hasegawa, C.S.\ Lim and N.\ Maru, hep-ph/0408028.
\reftitle{An Atempt to Solve the Hierarchy Problem Based on Gravity-Gauge-Higgs Unification Scenario}

\bibitem{Antoniadis1}
I.\ Antoniadis, K.\ Benakli and M.\ Quiros,
\jnl{\it New. J.\ Phys.}{3}{20}{2001}.
\reftitle{Finite Higgs mass without Supersymmetry}

\bibitem{Kawamura}
Y.~Kawamura, \jnl{\PTP}{103}{613}{2000}; 
\reftitle{Gauge Symmetry Reduction from the Extra Space $S^1/Z_2$}
\jnl{\PTP}{105}{999}{2001}.
\reftitle{Triplet-doublet Splitting, Proton Stability and Extra Dimension}

\bibitem{Hall1}
L.\ Hall and Y.\ Nomura, \jnl{\PRD}{64}{055003}{2001}; 
\reftitle{Gauge Unification in Higher Dimensions}
\jnl{\AP}{306}{132}{2003};
\reftitle{Grand Unification in Higher Dimensions}

R.\ Barbieri, L.\ Hall and Y.\ Nomura,
\jnl{\PRD}{66}{045025}{2002};
\reftitle{Softly Broken Supersymmetric Desert from Orbifold Compactification}
\jnl{\NPB}{624}{63}{2002};
\reftitle{Models of Scherk-Schwarz Symmetry Breaking in 5D: Classification and   Calculability}

A.\ Hebecker and J.\ March-Russell,
\jnl{\NPB}{625}{128}{2002};
\reftitle{The Structure of GUT Breaking by Orbifolding}

M.\ Quiros,  hep-ph/0302189.
\reftitle{New Ideas in Symmetry Breaking}

\bibitem{Lim1}
M.\ Kubo, C.S.\ Lim and H.\ Yamashita,
 \jnl{\MPLA}{17}{2249}{2002}.
\reftitle{The Hosotani Mechanism in Bulk Gauge Theories with an Orbifold Extra   Space $S^1/Z_2$}

\bibitem{gaugeHiggs2}
G.\ Dvali, S.\ Randjbar-Daemi and R.\ Tabbash, 
\jnl{\PRD}{65}{064021}{2002};
\reftitle{The Origin of Spontaneous Symmetry Breaking in Theories with Large Extra   Dimensions}

L.J.\ Hall, Y.\ Nomura and D.\ Smith,  \jnl{\NPB}{639}{307}{2002};
\reftitle{Gauge-Higgs Unification in Higher Dimensions}

L.\ Hall, H.\ Murayama, and Y.\ Nomura, 
   \jnl{\NPB}{645}{85}{2002};
\reftitle{Wilson Lines and Symmetry Breaking on Orbifolds}

G.\ Burdman and Y.\ Nomura, \jnl{\NPB}{656}{3}{2003}; 
\reftitle{Unification of Higgs and Gauge Fields in Five Dimensions}

C.\ Csaki, C.\ Grojean and H.\ Murayama, \jnl{\PRD}{67}{085012}{2003};
\reftitle{Standard Model Higgs From Higher Dimensional Gauge Fields}

C.A.\ Scrucca, M.\ Serone and L.\ Silverstrini, \jnl{\NPB}{669}{128}{2003}; 
\reftitle{Electroweak symmetry breaking and fermion masses from extra dimensions}

K.\ Choi et al., 
\jnl{\JHEP}{0402}{37}{2004};
\reftitle{Electroweak Symmetry Breaking in Supersymmetric Gauge-Higgs
Unification Models}

C.A.\ Scrucca, M.\ Serone, L.\ Silvestrini and A.\ Wulzer,
\jnl{\JHEP}{0402}{49}{2004}.
\reftitle{Gauge-Higgs Unification in Orbifold Models} 

\bibitem{Nandi1}
I.\ Gogoladze, Y.\ Mimura and S.\ Nandi, 
\jnl{\PLB}{562}{307}{2003};
\reftitle{Unification of Gauge, Higgs and Matter in Extra Dimensions}
\jnl{\PRL}{91}{141801}{2003};
\reftitle{Unity of Elementary Particles and Forces In Higher Dimensions}
\jnl{\PRD}{69}{075006}{2004}.
\reftitle{Model Building with Gauge-Yukawa Unification}

\bibitem{HHHK}
N.\ Haba, M.\ Harada, Y.\ Hosotani and Y.\ Kawamura, 
\jnl{\NPB}{657}{169}{2003};   
{\it Erratum}, {\it ibid.}  B{\bf 669} (2003) {381};
\reftitle{Dynamical Rearrangement of Gauge Symmetry on the Orbifold $S^1/Z_2$}

N.\ Haba,  Y.\ Hosotani and Y.\ Kawamura, 
\jnl{\PTP}{111}{265}{2004}.
\reftitle{Classification and dynamics of equivalence classes in SU(N) gauge theory   on the orbifold $S^1/Z_2$}

\bibitem{YH5}
Y.\ Hosotani,  in {\it ''Strong Coupling Gauge Theories and Effective Field
Theories"},  ed. M. Harada, Y. Kikukawa and K. Yamawaki (World Scientific
2003), p.\ 234. (hep-ph/0303066).
\reftitle{GUT on Orbifolds: Dynamical Rearrangement of Gauge Symmetry}

\bibitem{HHKY}
N.\ Haba,  Y.\ Hosotani,  Y.\ Kawamura and T.\ Yamashita, 
\jnl{\PRD}{70}{015010}{2004}.
\reftitle{Dynamical symmetry breaking in Gauge-Higgs unification on orbifold}

\bibitem{HNT}
Y.\ Hosotani, S.\ Noda and K.\ Takenaga,
\jnl{\PRD}{69}{125014}{2004}.
\reftitle{Dynamical Gauge Symmetry Breaking and Mass Generation on the Orbifold $T^2/Z_2$}

\bibitem{LeeHo}
J.E.\ Hetrick and C.L.\ Ho, \jnl{\PRD}{40}{4085}{1989};
\reftitle{Dynamical Symmetry Breaking from Toroidal Compactification}

C.C.\ Lee and C.L.\ Ho, \jnl{\PRD}{62}{085021}{2000}.
\reftitle{Recurrent dynamical symmetry breaking and restoration by Wilson lines at   finite densities on a torus}

\bibitem{YH3}
Y.\ Hosotani, hep-ph/0408012.
\reftitle{Dynamical Gauge-Higgs Unification}

\bibitem{Csaki1}
C.\ Csaki, C.\ Grojean, L.\ Pilo, and J.\ Terning, 
\jnl{\PRL}{92}{101802}{2004}.
\reftitle{Towards a Realistic Model of Higgsless Electroweak Symmetry Breaking}

\bibitem{Oda1}
K.\ Oda and A.\ Weiler, hep-ph/0410061.
\reftitle{Wilson Lines in Warped Space: Dynamical Symmetry Breaking and Restoration}

\bibitem{Takenaga1}
K.\ Takenaga,  \jnl{\PLB}{425}{114}{1998};
\reftitle{Supersymmetry Breaking through Boundary Conditions Associated with the   $U(1)_{R}$}
\jnl{\PRD}{58}{026004}{1998};
\reftitle{Softly Broken Supersymmetric Gauge Theories through Compactifications}
\jnl{\PRD}{64}{066001}{2001}; 
\reftitle{Dynamics of Nonintegrable Phases in Softly Broken Supersymmetric Gauge   Theory with Massless Adjoint Matter}
\jnl{\PRD}{66}{085009}{2002}.
\reftitle{Gauge Symmetry Breaking through the Hosotani Mechanism in Softly Broken   Supersymmetric QCD}

\bibitem{Takenaga2}
K.\ Takenaga, \jnl{\PLB}{570}{244}{2003};
\reftitle{Effect of Bare Mass on the Hosotani Mechanism}

M.\ Boz and N.K.\ Pak, hep-ph/0402238.
\reftitle{Effects of Fermion Masses and Twisting on Non-Integrable Phases on   Compact Extra Dimensions}


\end{thebibliography}
\end{document}